\documentclass[submitting]{nst}

\usepackage{subfigure,dcolumn}
\usepackage[T2A,T1]{fontenc}
\usepackage[russian,english]{babel}
\usepackage{hyperref}
\usepackage{ulem}
\usepackage{lineno}

\usepackage{listings}
\begin{document}

\title{The Energy Response of LaBr$_{3}$(Ce), LaBr$_{3}$(Ce,Sr) and NaI(Tl) Crystals for GECAM}\thanks{Supported by the National Key Research and Development Program (Grant Nos. 2022YFB3503600 and 2021YFA0718500) and the Strategic Priority Research Program of the Chinese Academy of Sciences (Grant No. XDA15360102), and National Natural Science Foundation of China (Grant Nos. 12273042 and 12075258)}

\author{Pei-Yi Feng}
\email[Corresponding author, ]{Pei-Yi Feng, Particle and Astrophysics Center, Institute of High Energy Physics, No. 19 (B), Yuquan Road, Laoshan Street, Shijingshan District, Beijing, China, 19151915020, fengpeiyi@ihep.ac.cn.}
\affiliation{Key Laboratory of Particle Astrophysics, Institute of High Energy Physics, Chinese Academy of Sciences, Beijing 100049, China}
\affiliation{University of Chinese Academy of Sciences, Chinese Academy of Sciences, Beijing 100049, China}
\author{Xi-Lei Sun}
\email[Corresponding author, ]{Xi-Lei Sun, Experimental Physics Center, Institute of High Energy Physics, No. 19 (B), Yuquan Road, Laoshan Street, Shijingshan District, Beijing, China, 13671137148, sunxl@ihep.ac.cn.}
\affiliation{State Key Laboratory of Particle Detection and Electronics, Institute of High Energy Physics, Chinese Academy of Sciences, Beijing 100049, China}
\author{Zheng-Hua An}
\email[Corresponding author, ]{Zheng-Hua An, Particle and Astrophysics Center, Institute of High Energy Physics, No. 19 (B), Yuquan Road, Laoshan Street, Shijingshan District, Beijing, China, 13661351124, anzh@ihep.ac.cn.}
\affiliation{Key Laboratory of Particle Astrophysics, Institute of High Energy Physics, Chinese Academy of Sciences, Beijing 100049, China}
\author{Yong Deng}
\affiliation{School of Nuclear Science and Technology, University of South China, Hengyang Hunan 421001, China}
\author{Cheng-Er Wang}
\affiliation{National Engineering Research Center for Rare Earth, Grirem Advanced Materials Co., Ltd. and General Research Institute for Nonferrous Metals, Beijing 100088, China}
\author{Huang Jiang}
\affiliation{State Key Laboratory of Particle Detection and Electronics, Institute of High Energy Physics, Chinese Academy of Sciences, Beijing 100049, China}
\author{Jun-Jie Li}
\affiliation{School of Nuclear Science and Technology, University of South China, Hengyang Hunan 421001, China}
\author{Da-Li Zhang}
\affiliation{Key Laboratory of Particle Astrophysics, Institute of High Energy Physics, Chinese Academy of Sciences, Beijing 100049, China}
\author{Xin-Qiao Li}
\affiliation{Key Laboratory of Particle Astrophysics, Institute of High Energy Physics, Chinese Academy of Sciences, Beijing 100049, China}
\author{Shao-Lin Xiong}
\affiliation{Key Laboratory of Particle Astrophysics, Institute of High Energy Physics, Chinese Academy of Sciences, Beijing 100049, China}
\author{Chao Zheng}
\affiliation{Key Laboratory of Particle Astrophysics, Institute of High Energy Physics, Chinese Academy of Sciences, Beijing 100049, China}
\affiliation{University of Chinese Academy of Sciences, Chinese Academy of Sciences, Beijing 100049, China}
\author{Ke Gong}
\affiliation{Key Laboratory of Particle Astrophysics, Institute of High Energy Physics, Chinese Academy of Sciences, Beijing 100049, China}
\author{Sheng Yang}
\affiliation{Key Laboratory of Particle Astrophysics, Institute of High Energy Physics, Chinese Academy of Sciences, Beijing 100049, China}
\author{Xiao-Jing Liu}
\affiliation{Key Laboratory of Particle Astrophysics, Institute of High Energy Physics, Chinese Academy of Sciences, Beijing 100049, China}
\author{Min Gao}
\affiliation{Key Laboratory of Particle Astrophysics, Institute of High Energy Physics, Chinese Academy of Sciences, Beijing 100049, China}
\author{Xiang-Yang Wen}
\affiliation{Key Laboratory of Particle Astrophysics, Institute of High Energy Physics, Chinese Academy of Sciences, Beijing 100049, China}
\author{Ya-Qing liu}
\affiliation{Key Laboratory of Particle Astrophysics, Institute of High Energy Physics, Chinese Academy of Sciences, Beijing 100049, China}
\author{Yan-Bing Xu}
\affiliation{Key Laboratory of Particle Astrophysics, Institute of High Energy Physics, Chinese Academy of Sciences, Beijing 100049, China}
\author{Xiao-Yun Zhao}
\affiliation{Key Laboratory of Particle Astrophysics, Institute of High Energy Physics, Chinese Academy of Sciences, Beijing 100049, China}
\author{Jia-Cong Liu}
\affiliation{Key Laboratory of Particle Astrophysics, Institute of High Energy Physics, Chinese Academy of Sciences, Beijing 100049, China}
\affiliation{University of Chinese Academy of Sciences, Chinese Academy of Sciences, Beijing 100049, China}
\author{Fan Zhang}
\affiliation{Key Laboratory of Particle Astrophysics, Institute of High Energy Physics, Chinese Academy of Sciences, Beijing 100049, China}
\author{Hong Lu}
\email[Corresponding author, ]{Hong Lu, Particle and Astrophysics Center, Institute of High Energy Physics, No. 19 (B), Yuquan Road, Laoshan Street, Shijingshan District, Beijing, China, 13681034963, luh@ihep.ac.cn.}
\affiliation{Key Laboratory of Particle Astrophysics, Institute of High Energy Physics, Chinese Academy of Sciences, Beijing 100049, China}

\begin{abstract}
The GECAM series of satellites utilize LaBr$_3$(Ce), LaBr$_3$(Ce,Sr), and NaI(Tl) crystals as sensitive materials for gamma-ray detectors (GRDs). To investigate the non-linearity in the detection of low-energy gamma rays and address errors in the E-C relationship calibration, comprehensive tests and comparative studies of the non-linearity of these three crystals were conducted using Compton electrons, radioactive sources, and mono-energetic X-rays. The non-linearity test results for Compton electrons and X-rays displayed substantial differences, with all three crystals showing higher non-linearity for X/$\gamma$-rays than for Compton electrons. Despite LaBr$_3$(Ce) and LaBr$_3$(Ce,Sr) crystals having higher absolute light yields, they exhibited a noticeable non-linear decrease in light yield, especially at energies below 400 keV. The NaI(Tl) crystal demonstrated "excess" light output in the 6–200 keV range, reaching a maximum "excess" of 9.2\% at 30 keV in X-ray testing and up to 15.5\% at 14 keV during Compton electron testing, indicating a significant advantage in the detection of low-energy gamma rays. Furthermore, this paper explores the underlying causes of the observed non-linearity in these crystals. This study not only elucidates the detector responses of GECAM, but also marks the inaugural comprehensive investigation into the non-linearity of domestically produced lanthanum bromide and sodium iodide crystals.
\end{abstract}

\maketitle

\nolinenumbers
\section{Introduction}

Recent years have witnessed groundbreaking advancements in various branches of astrophysics, such as gravitational waves, fast radio bursts, and cosmic rays, paving the way for a new "multi-messenger, multi-wavelength" era in astronomy\cite{2017Observation,2017Gravitational,2018GW170817,2017Multi,2018Insight}. These discoveries emphasize the importance of efficient detection methods for further understanding of high-energy astronomical phenomena. Transient gamma-ray sources, including gamma ray bursts and magnetar flares, play a vital role in shaping the landscape of astronomical research\cite{2015Localization,M2019A,2022Energetic,2018A,SSPMA-2020-0457,SSPMA-2019-0397}.

The Gravitational wave burst high-energy Electromagnetic Counterpart All-sky Monitor (GECAM) series, comprising satellites GECAM-A/B, GECAM-C and GECAM-D, was developed to monitor various high-energy electromagnetic events such as gamma-ray bursts and magnetar flares\cite{SSPMA-2019-0415,SSPMA-2020-0012,SSPMA-2019-0417,2022The,ZHANG2023168586}. These satellites employ gamma-ray detectors (GRDs) that utilize different scintillating crystals such as LaBr$_{3}$(Ce) and LaBr$_{3}$(Ce,Sr) for GECAM-A/B, and a combination of LaBr$_{3}$(Ce), LaBr$_{3}$(Ce,Sr), and NaI(Tl) for GECAM-C to validate new detector technologies. The fourth satellite, GECAM-D, uses NaI(Tl) crystals, and is scheduled to be launched in early 2024. The main characteristics of the GRDs are listed in Table~\ref{tab:characteristics of the GRD}.

\begin{table*}[!htb]
\caption{The main characteristics of the GRD in GECAM series.}
\label{tab:characteristics of the GRD}
\begin{tabular*}{16cm} {@{\extracolsep{\fill} } cccc}
\toprule
GRD Parameter & GECAM-A/B & GECAM-C & GECAM-D \\
\midrule
Type & LaBr$_3$(Ce); LaBr$_3$(Ce+Sr) & NaI(Tl); LaBr$_3$(Ce); LaBr$_3$(Ce+Sr) & NaI(Tl) \\
Quantity & 25 & 12 & 5\\
Area & 45.36 cm$^2$ & 45.36 cm$^2$ & 103.87 cm$^2$\\
Energy range & 8–2000 keV & 15–4000 keV & 20–1000 keV\\
Energy resolution & < 18\%@59.5 keV & < 18\%@59.5 keV & < 25\%@59.5 keV\\
Detection efficiency & > 50\%@8 keV & > 75\%@15 keV & > 60\%@20 keV\\
Deadtime & 4 $\mu$s & 4 $\mu$s & 4 $\mu$s\\
\bottomrule
\end{tabular*}
\end{table*}

\begin{figure}[!htb]
\includegraphics
  [width=0.5\hsize]
  {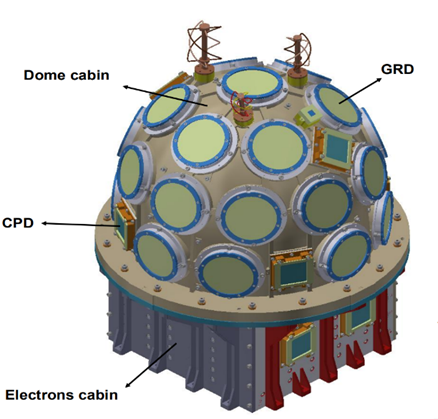}
\caption{The detector layout schematic of the GECAM-A/B payload. On each satellite, the detectors are designed with a modular approach, consisting of 25 Gamma-Ray Detector (GRD) modules and 8 Charged Particle Detector (CPD) modules.}
\label{fig:GECAM-A/B payload}
\end{figure}

GRDs serve as the primary detectors in the GECAM payload, and GECAM-A/B utilize an innovative solution employing LaBr$_{3}$ crystals coupled with silicon photomultiplier (SiPM) readout technology (Fig.~\ref{fig:GECAM-A/B payload})\cite{SSPMA-2019-0417,2022The,2022Electron,lu2022monte}. LaBr$_{3}$ crystals are advanced inorganic scintillators known for their high light output, excellent energy and timing resolutions, good energy linearity, and short light decay time. The SiPM, which replaces the conventional photomultiplier tube (PMT), offers advantages such as a simple and compact structure, ease of miniaturization, and efficient readout capability.

\begin{figure}[!htb]
\includegraphics
  [width=1\hsize]
  {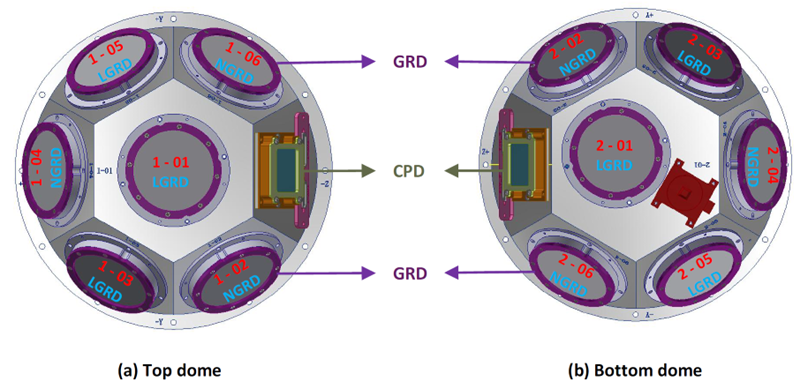}
\caption{The detector layout schematic of the GECAM-C payload. GECAM-C is composed of two detector domes: the top and bottom domes. Each dome is equipped with different types of GRDs, including NaI-based gamma-ray detectors (NGRDs) and LaBr$_3$-based gamma-ray detectors (LGRDs).}
\label{fig:GECAM-C payload}
\end{figure}

For GECAM-C (Fig.~\ref{fig:GECAM-C payload}), the GRDs employ both NaI(Tl) and LaBr$_{3}$ crystals, which are coupled to SiPM readout arrays\cite{ZHANG2023168586,Ground-calibration-GECAM-C}. The NaI(Tl) crystal is a high-performance, traditional inorganic scintillator with excellent luminescence properties that provides good resolution for both X-rays and gamma rays. Inorganic scintillators are widely used as the preferred choice for high-energy X/$\gamma$-ray detectors in space due to their versatility in shaping and sizing, stability, reliability, reasonable cost, inclusion of heavy elements, high density, and efficient detection capabilities for X/$\gamma$-rays.

The crystals used in the GECAM satellite series were produced by the Beijing Glass Research Institute. To optimize the performance of these detectors, it is vital to understand their energy responses. Consequently, we conducted an in-depth study involving X-ray, Compton electron, and gamma-ray tests on LaBr$_{3}$(Ce), LaBr$_{3}$(Ce,Sr), and NaI(Tl) crystals used in these satellites\cite{2023WW,2023HH,2023490}. Our findings indicated that the non-linearity of the three crystals varied when exposed to distinct excitation sources. The LaBr$_{3}$(Ce,Sr) crystal exhibited the strongest linear response to Compton electrons in the low-energy range, while the NaI(Tl) crystal demonstrated the best linear response to X-rays.

Consistent with previous publications on the non-linearity of iodide crystals\cite{moszynski2016energy,LIMKITJAROENPORN201815110}, domestically produced NaI(Tl) crystals exhibited an light yield "excess" phenomenon, indicating unexpected advantages in the detection of low-energy gamma rays. These insights not only contribute to a deeper understanding of the detector response of the GECAM series but also serve as invaluable information for evaluating the performance of these domestically produced scintillating crystals in the low-energy range of 3–400 keV\cite{yang2022response}. Manufacturers can refer to this paper to enhance their understanding of crystal non-linearity, potentially facilitating optimization and improvement in crystal growth processes and doping ratios. Furthermore, this study discusses the issue of non-linearity in crystals for low-energy gamma-ray detection, which holds substantial significance in addressing errors in detector calibration related to the energy-channel (E-C) relationship. Based on the content of this paper, we will delve deeper into the intrinsic resolution of crystals in our future work.


\section{Experimental Setup and Test Procedure}

\subsection{The Wide-Angle Compton Coincidence Technology}

\begin{figure}[!htb]
\includegraphics
  [width=0.7\hsize]
  {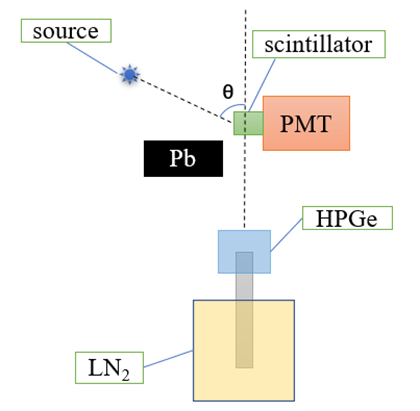}
\caption{Schematic diagram of experimental setup for obtaining Compton electrons with a wide energy range.}
\label{fig:WACC setup}
\end{figure}

    Figure~\ref{fig:WACC setup} shows the Wide-Angle Compton Coincidence (WACC) experimental setup, which primarily comprises a radioactive source, an HPGe detector, the scintillation detector under examination, and a subsequent data acquisition system\cite{dong2023development,2017Intrinsic,LIMKITJAROENPORN201815110,2009A,2012Non,zhang2022transition}. LaBr$_{3}$(Ce), LaBr$_{3}$(Ce,Sr), and NaI(Tl) cylindrical samples with diameters and height of 25.4 mm were selected for this study. Silicone oil was used to couple the encapsulated crystals to PMTs, which were R6233-100 models produced by Hamamatsu Photonics, Japan\cite{Hamamatsu,han2023csi}. According to the user manual of the BE2020 planar germanium spectrometer manufactured by Canberra, the HPGe crystal had a thickness of 20 mm and volume of 40,000 mm$^{3}$, allowing for a detection energy range of 3 keV to 3 MeV\cite{Canberra,BEGe}. Based on the experimental data, the energy resolution (represented by the full width at half maximum, i.e., FWHM) of the HPGe detector used in this study was determined to be 1.58 keV ($^{60}$Co, 1.33 MeV) and 1.15 keV ($^{137}$Cs, 662 keV).

\begin{figure*}[!htb]
\includegraphics
  [width=0.9\hsize]
  {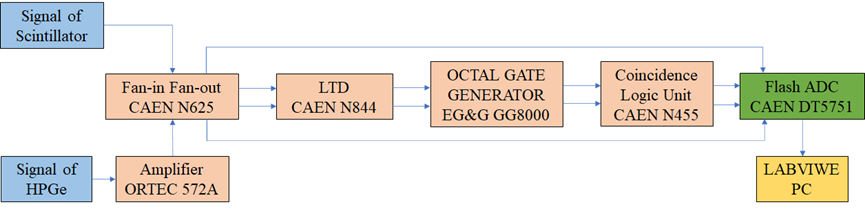}
\caption{Diagram of the data acquisition system for coincidence events.}
\label{fig:data acquisition system}
\end{figure*}

\begin{figure}[!htb]
    \includegraphics
      [width=0.9\hsize]
      {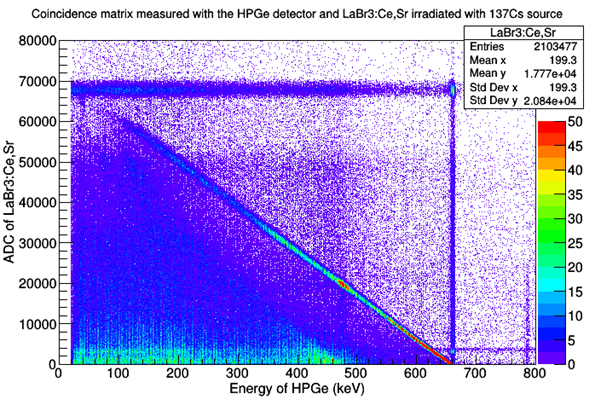}(a)
    \includegraphics
      [width=0.9\hsize]
      {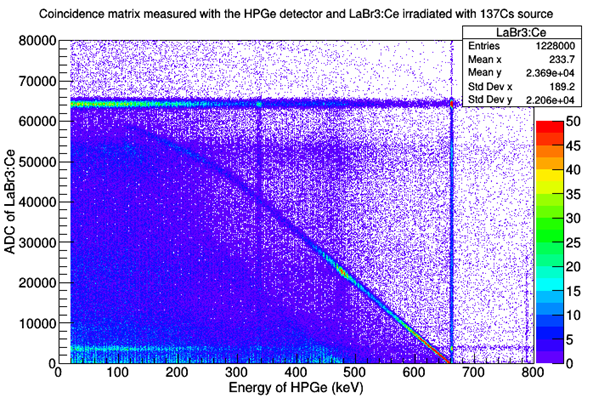}(b)
    \includegraphics
      [width=0.9\hsize]
      {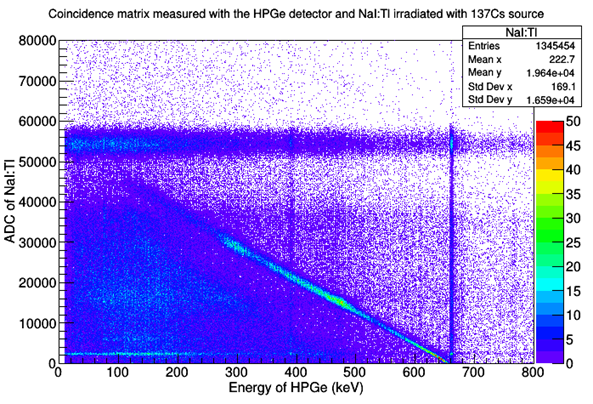}(c)
\caption{Two-dimensional spectrum of coincidence events in both HPGe and LaBr$_3$(Ce,Sr) (a), LaBr$_3$(Ce) (b), NaI(Tl) (c) with $^{137}$Cs source, i.e., coincidence matrix.}
\label{fig:2D coincidence matrix}
\end{figure}

The experiment involved placing a $^{137}$Cs radioactive source at a quarter-circle position around the center of the crystal, with a distance of 13 cm between them. Gamma photons were emitted by the radioactive source via radioactive decay and underwent Compton scattering when they struck the crystal. Consequently, Compton electrons were generated and absorbed in the crystal, whereas some scattered photons escaped from the crystal and were absorbed by the nearby HPGe detector. The distance between the tested crystal and HPGe detector was maintained at approximately 15 cm\cite{dong2023development}. Lead blocks were positioned between the $^{137}$Cs source and HPGe detector to provide shielding and minimize the incidence of primary gamma photons directly irradiating the HPGe detector. Coincidence events across a broad energy range could be obtained by adjusting the position of the radioactive source and varying the angle between the source, crystal, and HPGe detector.

A desktop waveform acquisition device with 10 bit @ 2 GS/s (interleaved) or 1 GS/s, the DT5751 digitizer\cite{LI2014Particle}, was utilized in this experiment to collect signals from the crystal and HPGe detectors (Fig.~\ref{fig:data acquisition system}). The HPGe detector signal operating at +3500 V underwent shaping and filtering via an ORTEC 572A amplifier before being sent to channel-0 of the DT5751. The output signal from the PMT anode operating at +1300 V was routed to channel-1 of the DT5751 after the photoelectrons were multiplied by the dynodes\cite{liu2023toward}. The signals from the crystal and HPGe detectors successively underwent low-threshold discrimination, delay stretching, and logical coincidence. The resulting coincidence output signal was utilized as an external trigger for the DT5751. The DT5751 records the corresponding coincidence events and generates two data files when triggered externally. The secondary particles produced by Compton scattering were absorbed by the two detectors in a specific order in the time sequence. For a "true coincidence event", the waveform signal corresponding to the crystal appeared before the HPGe detector.

Figure~\ref{fig:2D coincidence matrix} displays the coincidence matrix that represents  all collected event data. In Fig.~\ref{fig:2D coincidence matrix} (a), the horizontal axis represents the energy deposited in the HPGe detector, whereas the vertical axis corresponds to the energy deposited in the LaBr$_{3}$(Ce,Sr) crystal. As stated in Compton scattering formula (Equation~\ref{eq:Compton scattering})\cite{LIMKITJAROENPORN201815110}, with an increase in the incident angle, $\theta$, of the gamma photon, the energy of the Compton electrons in the LaBr$_{3}$(Ce,Sr) crystal is also increased.

\begin{equation}\label{eq:Compton scattering}
E_e = E_{\gamma} - E_{\gamma}^{'} = \frac{E_{\gamma}}{1 + \frac{m_{e}c^2}{E_{\gamma}(1 - cos\theta)}} .
\end{equation}

where E$_\gamma$ is the energy of the gamma ray radiated from the source, E$_e$ is the energy of the Compton electron, E$_\gamma^{'}$ is the energy of the scattered photon, $\theta$ is the Compton scattering angle, and m$_e$c$^2$ is the remaining mass energy of the electron. Five scattering angles of $\theta$ were chosen during the experiment to obtain data over a broad energy range.

The diagonal points in Fig.~\ref{fig:2D coincidence matrix} denote the "true coincidence events" of interest in this study. Each point corresponds to a specific scattering angle, and the combined deposited energies in both the crystal and the HPGe detector are constant of 661.6 keV. The uneven "spread" among the diagonal at different energy levels is due to the diverse energy resolutions of the crystal for Compton electrons. Furthermore, the non-linear response of the crystal determines the "linearity" of the diagonal. Coincidence matrix analysis enabled the extraction of the energy resolution and non-linear response of the crystal for Compton electrons.

Figure~\ref{fig:2D coincidence matrix} displays the horizontal and vertical lines indicating accidental coincidence events detected simultaneously by both detectors. The horizontal line represents the finite resolution of the crystal and the vertical line represents the excellent resolution of the HPGe detector. The other points on the graph denote events in which the partial energy was deposited in the detector or where detection occurred after scattering through the surrounding materials.

The WACC method accurately measures the energy response and resolution of the crystal detector to Compton electrons. Before the Compton experiments, it was necessary to calibrate the E–C relationship of the HPGe detector, which can be obtained from the energy spectra of multiple radioactive sources or directly using the vertical lines in the coincidence matrix.

The HPGe detector offers an outstanding energy resolution, making it an excellent standard detector. The energy deposited in the crystal can be calculated by subtracting the scattered photon energy in the HPGe detector from the known gamma-ray source energy. In actual data processing, the cut width of the HPGe energy axis must be determined based on the Compton scattering event statistics. Within this range, the central value is considered to be the deposited energy in the HPGe detector, and Equation~\ref{eq:Scin-energy} is used to calculate the energy deposited in the crystal. In this study, cutting was performed on the HPGe energy axis and projection onto the crystal axis.

\begin{equation}\label{eq:Scin-energy}
<E_{scin}> = E_{\gamma} - <E_{HPGe}> .
\end{equation}

where E$_\gamma$ is the known gamma-ray source energy, <E$_{HPGe}$> is the deposited energy in the HPGe detector, and <E$_{scin}$> is the deposited energy in the crystal.

\begin{figure}[!htb]
\includegraphics
  [width=0.98\hsize]
  {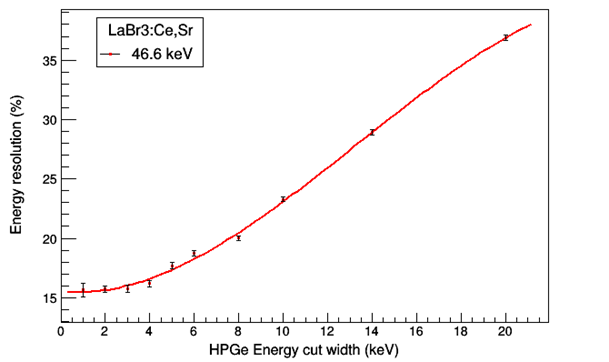}
\caption{Resolution of Compton electrons in LaBr$_3$(Ce,Sr) versus cut width in HPGe for a $^{137}$Cs source.}
\label{fig:cut width}
\end{figure}

To understand the effect of the cut width or energy window width, we measured the energy resolution of the LaBr$_{3}$(Ce,Sr) crystal to 46.6 keV Compton electrons for different cut widths. The outcome indicated that the energy resolution remained reasonably stable until a cut width of 4 keV was reached (Fig.~\ref{fig:cut width}). Wider cut-widths led to a broadened scattering angle range of relevant valid events and an increase in the FWHM of the Compton electron spectrum. Subsequently, the resolution deteriorated. The energy resolution of the HPGe detector was within the range of 1–2 keV, which must be considered when determining a reasonable cut width. It is also essential to ensure a sufficient number of events. Therefore, a cutoff width of 4 keV was used when the energy deposited in the HPGe detector was less than 615 keV. When the deposited energy fell at 615–661.6 keV, a cut width of 2 keV was selected.

\begin{figure}[!htb]
    \includegraphics
      [width=0.88\hsize]
      {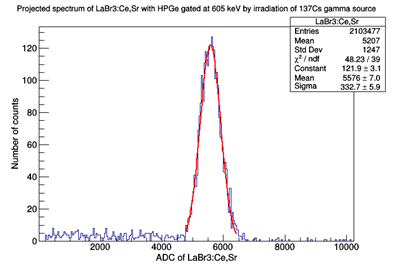}(a)
    \includegraphics
      [width=0.88\hsize]
      {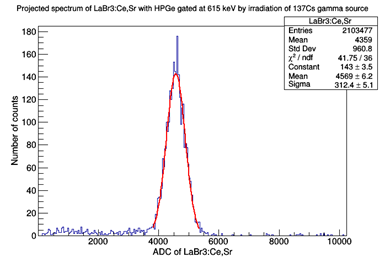}(b)
    \includegraphics
      [width=0.88\hsize]
      {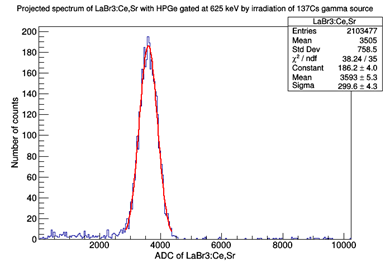}(c)
    \includegraphics
      [width=0.88\hsize]
      {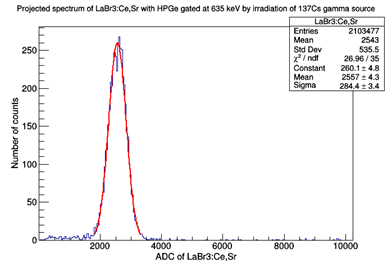}(d)
\caption{Projected spectra of LaBr$_3$(Ce,Sr) crystal with HPGe energy gated at 605 keV (a), 615 keV (b), 625 keV (c) and 635 keV (d), respectively, i.e., the 56.6 keV (a), 46.6 keV (b), 36,6 keV (c) and 26.6 keV (d) Compton electron energy spectrum.}
\label{fig:electron spectrum}
\end{figure}

Multiple truncations of the HPGe energy axis were performed to obtain the spectra of the crystal for various Compton electron energies. Figure~\ref{fig:electron spectrum} illustrates an example of this approach, in which an event data range of 614–616 keV was considered at an HPGe energy of 615 keV with a cut width of 2 keV to produce the Compton electron spectrum (Fig.~\ref{fig:electron spectrum} (b)). A Gaussian-shaped, single-energy electron peak was visible and fitting it with a Gaussian function returned an energy resolution of 15.81 ± 0.25\% for 46.6 keV Compton electrons in the LaBr$_{3}$(Ce,Sr) crystal.

When incident particles deposit energy in a crystal, it causes excitation of atoms or molecules, leading to the emission of scintillation photons with wavelengths similar to those of visible light\cite{qian2021simulation}. The light yield, defined as the number of scintillation photons per unit of energy deposited in the crystal, is described by Equation~\ref{eq:light yield}.

\begin{equation}\label{eq:light yield}
S = \frac{ADC}{ADC_{spe} \cdot E} .
\end{equation}

where S describes the light yield of the crystal, ADC represents the spectrum's peak position after deducting the baseline, E is the deposited energy in the crystal, and ADC$_{spe}$ = 8.0321 channels denotes the single-photoelectron response of the Hamamatsu R6233-100 PMT at a high voltage of $+$1300 V. The response was calibrated using the LED-triggered charge method\cite{2018Consistency,li2023performance}.

\subsection{Measured by Radioactive Sources}

We employed radioactive sources of $^{133}$Ba, $^{137}$Cs, $^{241}$Am, $^{152}$Eu, and $^{207}$Bi across a range of $\gamma$-ray energies from 30.85 keV to 1063.7 keV to investigate the gamma-ray responses. The tested crystal was coupled to a Hamamatsu R6233-100 PMT via silicon oil, and the digitizer DT5751 acquired the signal waveforms in self-triggering mode. ROOT, a data analysis framework conceived by the European Organization for Nuclear Research (CERN)\cite{wu2021root}, had been used for analysing the experimental data in this study, including baseline subtraction, fitting of the full-energy peak, and analysis of the peak position and FWHM.

\subsection{Single Energy X-ray Measurements Using the Hard X-ray Calibration Facility}

\begin{figure*}[!htb]
\includegraphics
  [width=0.75\hsize]
  {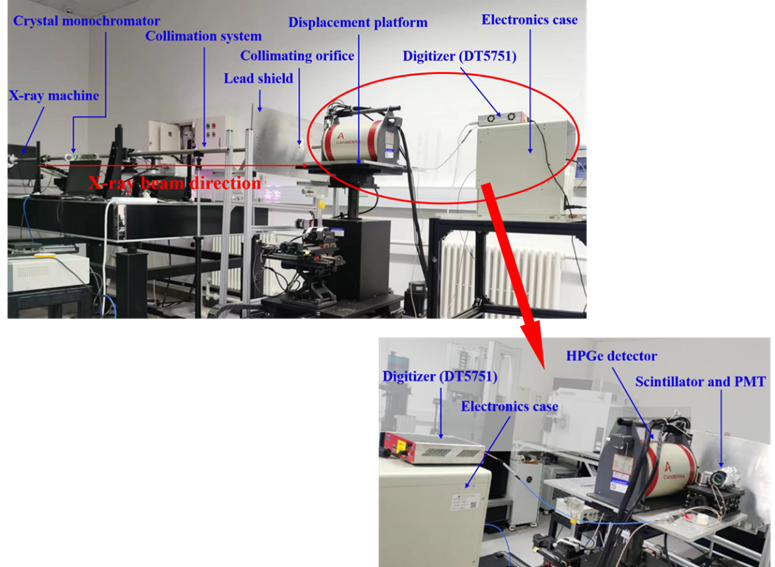}
\caption{The Hard X-ray Calibration Facility. Both the HPGe and the crystal detectors were placed on a displacement platform and kept on the same horizontal line.}
\label{fig:HXCF}
\end{figure*}

We employed two sets of hard X-ray calibration facilities (HXCF, Fig.~\ref{fig:HXCF}) established by the National Institute of Metrology (NIM) in Beijing Changping, China\cite{2021The,2019The,2022Research} to investigate the energy responses of these three crystals to X-rays in the range of 8–120 keV. The HXCF, which plays a substantial role in the calibration of gamma-ray detectors on GECAM, CubeSats and SVOM satellites\cite{2021Ground,wen2021compact,2021Calibration}, was first built for the high-energy telescope of HXMT as a calibration facility\cite{2019Ground} and comprises four primary components: an X-ray generator, monochromator, collimator, and standard detector. To shielding stray light from the X-ray generator, the collimator features apertures of various sizes at the entrance and exit. A low-energy HPGe detector from Canberra Industries was used as a standard detector. Before testing, we calibrated the HPGe detector for energy linearity, energy resolution, and detection efficiency using various standard radioactive sources\cite{2016LEGe}.

\begin{figure*}[!htb]
\includegraphics
  [width=0.8\hsize]
  {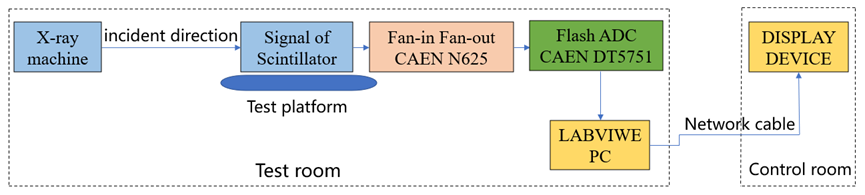}
\caption{The data acquisition system diagram for single-energy X-ray detection.}
\label{fig:single-energy X-ray detection}
\end{figure*}

\begin{figure}[!htb]
\includegraphics
  [width=1.\hsize]
  {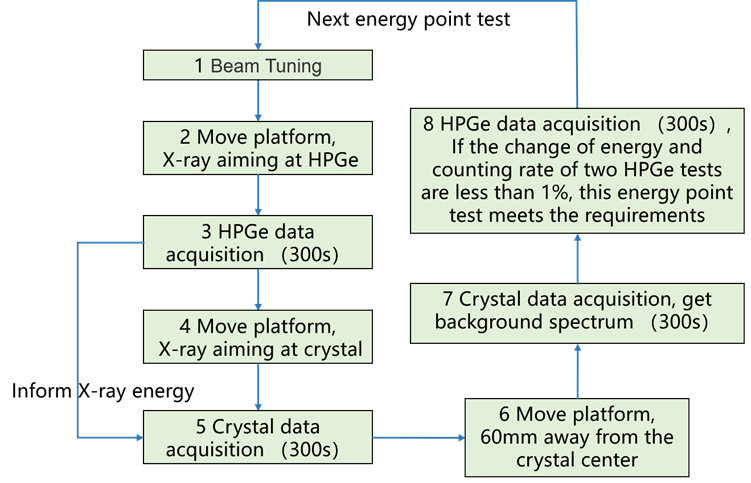}
\caption{The test procedure for the non-linear response of the crystals to X-rays.}
\label{fig:test procedure}
\end{figure}

The entire set of testing equipment, including the data-acquisition system, was placed inside an X-ray testing chamber (Fig.~\ref{fig:single-energy X-ray detection}) and remotely controlled for data retrieval in the control room. The energy and flux of the X-rays were determined by the HPGe detector and the testing procedures are shown in Fig.~\ref{fig:test procedure}. We utilized GENIE 2000, which is a spectroscopic data acquisition and analysis software, to record the spectral data from the HPGe detector. The crystal detector was coupled to a PMT (Hamamatsu Model CR160) using silicone oil. The signals from the crystal detector were collected using a digitizer (DT5751) and analyzed using computer software for the corresponding spectra.

In this study, the range of X-ray testing was 8–120 keV, with fine measurement of the crystal's absorption edge at a step size of 0.1 keV. The performance of the crystal detector gradually changed with increasing X-ray energy, allowing for a reduced number of test energy points. Owing to the testing at room temperature (22–23 ℃), the detector noise was slightly higher, limiting the starting test energy points to the range of 8–10 keV. For the two LaBr$_3$ crystals, the PMTs coupled to them operated at $-$800 V, while the NaI(Tl) crystal was at $-$1000 V.

\section{Results and Discussion}

\subsection{Light Yield Non-linearity to Compton Electrons}

\begin{figure}[!htb]
\includegraphics
  [width=1.\hsize]
  {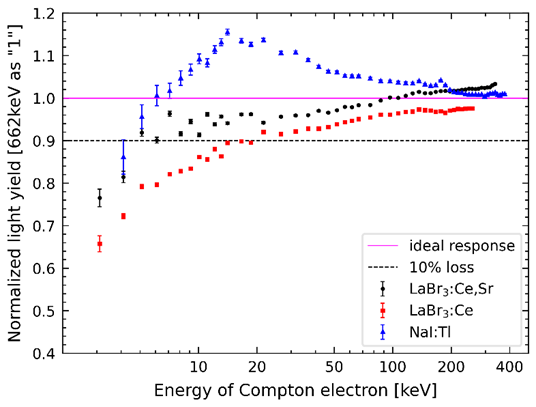}
\caption{Comparison of the light yield non-linearity of LaBr$_3$(Ce), LaBr$_3$(Ce,Sr), and NaI(Tl) crystals for 3–400 keV Compton electrons.}
\label{fig:electrons light yield}
\end{figure}

The light yields of LaBr$_3$(Ce), LaBr$_3$(Ce,Sr), and NaI(Tl) crystals were normalized to "1" at 662 keV energy. Figure~\ref{fig:electrons light yield} depicts the light output non-linearity of the LaBr$_3$(Ce), LaBr$_3$(Ce,Sr), and NaI(Tl) crystals to Compton electrons within the energy range of 3–400 keV. To better quantify the non-linearity of these crystals, we introduced a metric known as the "Non-linearity Standard Deviation" (NLSD), denoted by Equation~\ref{eq:metric}, where x$_i$ represents the relative light yield at each energy point. The NLSD values for the LaBr$_3$(Ce), LaBr$_3$(Ce,Sr), and NaI(Tl) crystals were calculated to be 0.11, 0.03, and 0.06, respectively. The larger the NLSD value, the more significant the crystal non-linearity.

\begin{equation}\label{eq:metric}
NLSD = \sqrt{\frac{1}{n}\sum_{i=1}^{n} (x_i - 1)^2}  \ \ \ \ (n =1, 2, 3...) .
\end{equation}

For both types of LaBr$_3$ crystals, as the energy of the Compton electrons decreased, the non-linearity of the light yield gradually increased. Within the measured electron energy range, the LaBr$_3$(Ce,Sr) crystal exhibited better linearity than the LaBr$_3$(Ce) crystal, particularly at energies below 20 keV. We hypothesize that the doping of Sr$^{2+}$ may have improved the internal energy transfer mechanism within the LaBr$_3$(Ce,Sr) crystal, enhancing energy transfer efficiency in the low-energy region, thereby ameliorating non-linearity. Both crystals exhibited a 10\% "defect" light output at approximately 5 keV and 20 keV, respectively. The minimum measurable energy point using WACC was 3.1 keV, at which the LaBr$_3$(Ce,Sr) crystal exhibited approximately 24\% "defect", whereas the LaBr$_3$(Ce) crystal reached a 35\% "defect". This experiment validated the electron non-linearity simulation results presented by Zheng Chao et al.\cite{2022Electron}, while also affirming the accuracy and rationality of both the model and experimental work conducted by the GECAM research team.

In contrast to the two LaBr$_3$ crystals, the NaI(Tl) crystal did not exhibit a monotonic "defect" luminescence phenomenon as the energy of Compton electrons decreased. At approximately 14 keV, the NaI(Tl) crystal reached its maximum light yield, exhibiting approximately 15.5\% "excess" light output. Beyond 14 keV, the light yield gradually decreased as the energy increased. Conversely, as the energy decreased below 14 keV, the light yield decreased. The lowest test energy point was 4.1 keV, at which the NaI(Tl) crystal demonstrated a luminosity non-linearity of approximately 14\% "defect".

\subsection{The Absolute Light Yield of Crystals}

The three crystals were irradiated using multiple radioactive sources to obtain the energy spectra of each crystal for different sources. The single-photoelectron responses of the Hamamatsu R6233-100 PMT used in the measurements were calibrated using the LED-triggered charge method at various voltages, enabling the calculation of the absolute light yields of these three crystals. The absolute light yields and energy resolutions of the tested samples at 661.6 keV are listed in Table~\ref{tab:absolute light yield}.

\begin{table}[!htb]
\caption{The absolute light yield and energy resolution of crystals for the 661.6 keV gamma rays.}
\label{tab:absolute light yield}
\begin{tabular*}{8.5cm} {@{\extracolsep{\fill} } cccc}
\toprule
Crystal Type & Size & Light Yield(ph/MeV) & Energy Resolution(\%) \\
\midrule
LaBr$_3$(Ce) & 1"×1" & 74196 & 3.00$\pm$0.02 \\
LaBr$_3$(Ce,Sr) & 1"×1" & 63939 & 3.03$\pm$0.01 \\
NaI(Tl) & 1"×1" & 45445 & 7.18$\pm$0.07 \\
\bottomrule
\end{tabular*}
\end{table}

\subsection{Energy Resolution}

\begin{figure}[!htb]
\includegraphics
  [width=1.\hsize]
  {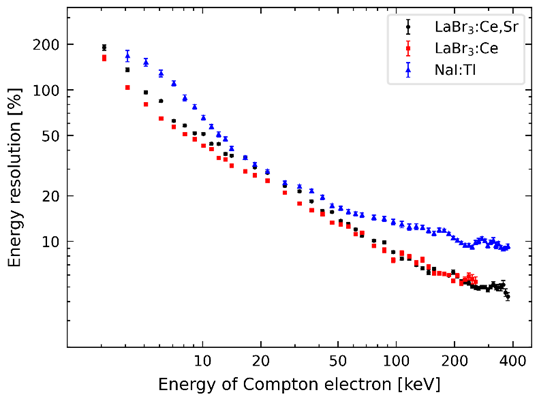}
\caption{Comparison of the energy resolution of LaBr$_3$(Ce), LaBr$_3$(Ce,Sr), and NaI(Tl) crystals for 3–400 keV Compton electrons.}
\label{fig:electrons resolution}
\end{figure}

Figure~\ref{fig:electrons resolution} illustrates the energy resolution of the LaBr$_3$(Ce), LaBr$_3$(Ce,Sr), and NaI(Tl) crystals for Compton electrons in the 3–400 keV range. The energy resolution of the NaI(Tl) crystal was comparable to that of the LaBr$_3$ crystals at 16–30 keV (Fig.~\ref{fig:electrons resolution}).

\begin{figure}[!htb]
\includegraphics
  [width=1.\hsize]
  {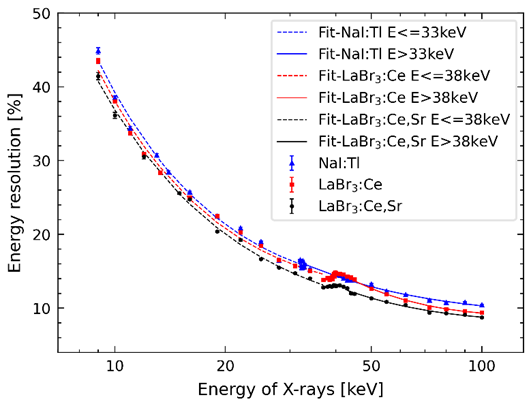}
\caption{The energy resolution of LaBr$_3$(Ce,Sr), LaBr$_3$(Ce), and NaI(Tl) crystals varies with X-ray energy.}
\label{fig:X-ray resolution}
\end{figure}

The energy resolution of the crystals was expressed using the FWHM of the X-ray full-energy peak. Figure~\ref{fig:X-ray resolution} shows the energy resolution of LaBr$_3$(Ce,Sr), LaBr$_3$(Ce), and NaI(Tl) crystals for X-rays in the 8–100 keV range as measured by HXCF. The LaBr$_3$(Ce,Sr) crystal exhibited the best energy resolution within this energy range. At 100 keV, the resolution of the LaBr$_3$(Ce,Sr) crystal was 8.74 ± 0.0681\%, while the LaBr$_3$(Ce) and NaI(Tl) crystals had resolutions of 9.41 ± 0.0976\% and 10.39 ± 0.1168\%, respectively. Furthermore, a slight degradation in the energy resolution of no more than 1\% was observed near the binding energy of the K-shell electrons.

\subsection{Comparison of the X/$\gamma$-Ray and Compton Electron Responses}

All data in this study were standardized by setting the full-energy peak response of 662 keV gamma rays from a $^{137}$Cs source as the normalization factor. The non-linearity of the LaBr$_3$(Ce,Sr) crystal's light yield for Compton electrons and gamma rays in the 3–1000 keV range is shown in Fig.~\ref{fig:LaBr3(Ce,Sr)}. Notably, the response of the Compton electrons exhibited excellent linearity at approximately 70 keV, with a non-linearity of less than 2\%. However, a "deficiency" in light output occurred when the energy of Compton electrons was below 70 keV, while substantial non-linearity was observed in the response to gamma rays below approximately 200 keV.

\begin{figure}[!htb]
\includegraphics
  [width=1.\hsize]
  {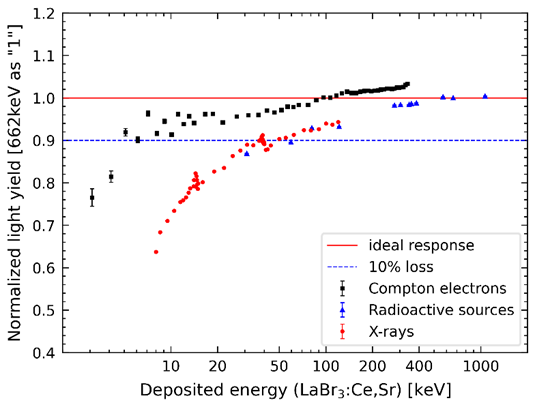}
\caption{The light yield non-linearity of LaBr$_3$(Ce,Sr) crystal for Compton electrons, gamma rays and X-rays.}
\label{fig:LaBr3(Ce,Sr)}
\end{figure}

A more detailed test was conducted on the photon response below 120 keV using HXCF. Figure~\ref{fig:LaBr3(Ce,Sr)} presents the non-linear light yield response curve of the LaBr$_3$(Ce,Sr) crystal to X-rays in the energy range 8–120 keV. As the error bars are similar in size to the data point symbols, they are not visible in the figure. Ideally, the relative light yield should be "1" at all energy points. However, this was not the case and varying degrees of light-yield deficiencies were observed within the tested energy range. Below 40 keV, the LaBr$_3$(Ce,Sr) crystal exhibited substantial non-linearity in the relative light yield response to X-rays, with the non-linearity exceeding 10\%. As the energy decreased, the slope of the curve increased, reaching a non-linearity of 36\% at 8 keV. When the X-ray energy exceeded 40 keV, the non-linear curve approached the ideal state and the slope became milder, indicating that the fluctuation in the number of photons generated per unit energy absorbed by the LaBr$_3$(Ce,Sr) crystal was small within the energy range of 40–120 keV. The LaBr$_3$(Ce,Sr) crystal exhibited absorption edges at 13–15 keV and 38–40 keV, and a slight reduction in the relative light yield was observed within these two energy intervals.

\begin{figure}[!htb]
\includegraphics
  [width=1.\hsize]
  {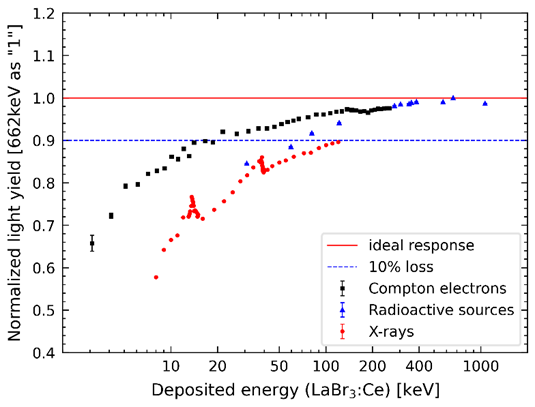}
\caption{The light yield non-linearity of LaBr$_3$(Ce) crystal for Compton electrons, gamma rays, and X-rays.}
\label{fig:LaBr3(Ce)}
\end{figure}

The NLSD values for testing the LaBr$_3$(Ce,Sr) crystal with X-rays and Compton electrons were 0.17 and 0.03, respectively. The light output of the LaBr$_3$(Ce,Sr) crystal exhibited greater non-linearity in response to X-rays than to Compton electrons. This can be attributed to the different mechanisms by which these particles interact with atoms in matter. For X/$\gamma$-rays ranging from a few keV to several hundred keV, there are two possible interaction processes with the crystal: (1) a direct photoelectric cascade sequence or (2) a Compton scattering followed by photoelectric cascade sequence. These processes generate several primary electrons (e.g., Compton electrons and primary photoelectrons) and multiple secondary electrons (e.g., Auger electrons and secondary photoelectrons), with the final light emission being the sum of the contributions from secondary electrons with different energies. Notably, these electrons are products of the interaction between the incident photons and matter, and their energies cannot exceed those of the incident particles. Therefore, the light output induced by the photons in the LaBr$_3$(Ce,Sr) crystal is always lower than that caused by Compton electrons with equivalent energies.

We conducted detailed testing of the LaBr$_3$(Ce) crystal using the same experimental procedures and data processing methods. Figure~\ref{fig:LaBr3(Ce)} illustrates the non-linear light yield response of the LaBr$_3$(Ce) crystal to Compton electrons and gamma rays across the energy range of 3–1000 keV. The non-linearity curves tend to be flat, and the results are similar for Compton electrons and gamma rays when the energy is above 200 keV, but substantial differences are observed below 200 keV. As the energy decreased, the LaBr$_3$(Ce) crystal exhibited a lower response to the full-energy peak of gamma-rays than to Compton electrons of the same energy. This finding is consistent with the test results for the LaBr$_3$(Ce,Sr) crystal, indicating that the manner in which the particles interact with matter directly affects the light output of the crystal. For gamma rays in the energy range of several hundred kiloelectronvolts, Compton scattering is most likely the initial interaction, and most gamma rays require multiple interactions for full absorption. The high-energy primary and secondary electrons resulting from these interactions exhibited good linearity in their response, reflecting the excellent linearity of the response to high-energy gamma rays.

\begin{figure}[!htb]
\includegraphics
  [width=1.\hsize]
  {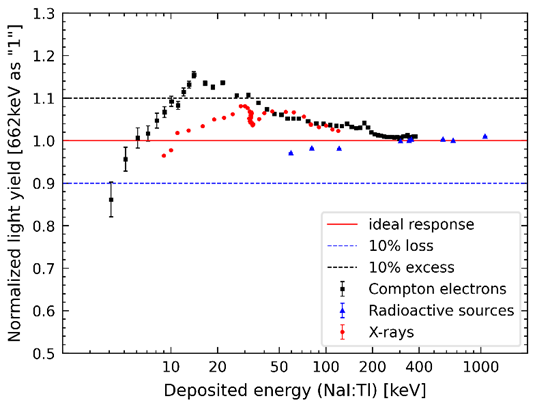}
\caption{The light yield non-linearity of NaI(Tl) crystal for Compton electrons, gamma rays and X-rays.}
\label{fig:NaI(Tl)}
\end{figure}

Figure~\ref{fig:LaBr3(Ce)} shows the non-linearity curve of the LaBr$_3$(Ce) crystal to X-rays in the energy range of 8–120 keV. Compared to the LaBr$_3$(Ce,Sr) crystal, this response curve deviated more markedly from the ideal state, and almost all the measured energy points exhibited scintillation responses below 90\%. The light output sharply decreased near the K-shell binding energies (13–15 keV and 38–40 keV) of Br and La, leading to a greater non-linearity of the LaBr$_3$(Ce) crystal response curve to X-rays. Data points below 28 keV exhibited non-linearity greater than 20\%, and the light output at 8 keV was only 58\% of the ideal state.

As X-ray energy decreased, induced secondary electron energies in the crystal decreased, thereby resulting in more significant light "defects". The NLSD values for testing the LaBr$_3$(Ce) crystal with X-rays and Compton electrons were 0.22 and 0.11, respectively. LaBr$_3$(Ce) crystal exhibited greater non-linearity to X-rays and Compton electrons than LaBr$_3$(Ce,Sr) crystal, particularly at energies below 100 keV. This may be attributed to the doping process, indicating that doping with Sr$^{2+}$ ions can improve the non-linearity of the LaBr$_3$ crystals.

To understand the differences in non-linearity among the different crystal types better, the NaI(Tl) crystal was chosen as the third test subject in this study (Fig.~\ref{fig:NaI(Tl)}). Unlike the two LaBr$_3$ crystals, the NaI(Tl) crystal exhibited a pronounced "excess" response to Compton electrons in the energy range of 8–80 keV, with a non-linearity exceeding 4\%. At electron energies lower than 6 keV, the crystal displayed slight "defects" in light output, while above 80 keV, the curve tended to flatten, indicating a good linear response of this sodium iodide compound to high-energy electrons.

Figure~\ref{fig:NaI(Tl)} also shows the non-linearity of the light yield of the NaI (Tl) crystal to X-rays in the energy range 8–120 keV. Compared to the response to Compton electrons, the X-ray test results exhibited a similar trend, with NLSD values of 0.06 for both. However, there were differences in the curve slopes. Direct photoelectric interactions with matter are most likely to occur for photons in the tens-of-keV range. Assuming this photoelectric absorption occurs with iodine K-shell electrons (with a probability of 83\% when the photon energy is greater than 33.17 keV), the resulting photoelectrons have energy falling within the range with substantial "excess" light output. The total light emission induced by all the secondary electrons generated from the photons exceeded that caused by Compton electrons with equivalent energies. Therefore, when the energy was within the range of 40–70 keV, NaI(Tl) crystal exhibited a higher relative light output to X-rays, producing a greater number of photons per unit X-ray-deposited energy compared to the case of Compton electron incidence.

The response of the NaI (Tl) crystal to X-rays was similar to that of Compton electrons at approximately 33 keV. This is related to the binding energy (33.17 keV) of the iodine K-shell electrons, as photons with energies lower than this energy cannot excite K-shell electrons from the iodine atoms. Almost all the photon energy was transferred to electrons, and only a small fraction of low-energy photons interacted with an iodine L-shell electron (with a binding energy of 5.19 keV) to produce lower-energy X-rays through the photoelectric effect.

Within the measured X-ray energy range, the NaI(Tl) crystal exhibited varying degrees of "excess" light output, which also can be explained by the photoelectric effect cascade sequence. In Fig.~\ref{fig:NaI(Tl)}, the low-energy electron response showed an "excess" and reached its maximum value at ~14 keV. Therefore, when photons undergo a series of interactions to produce multiple low-energy secondary electrons, a "burst" phenomenon occurs in the light output. This also explains why the photon response reached a maximum value at approximately 30 keV instead of 14 keV. As the incident photon energy increased, the light output gradually decreased but remained above 100\%. This is because of the more complex distribution of secondary electron energies, resulting in a large number of secondary electrons with energies lower than 6 keV. The electron response below 6 keV exhibited a "deficient" luminous response, which formed a so-called "compensation" effect with the "excess" phenomenon observed in the electron response in the tens of keV range.

\section{Conclusion}

We employed the WACC technique and HXCF/radioactive sources to compare the energy responses of domestically produced LaBr$_3$(Ce), LaBr$_3$(Ce, Sr), and NaI(Tl) crystals to Compton electrons and X/$\gamma$-rays. The NLSD values obtained through X-ray testing for LaBr$_3$(Ce), LaBr$_3$(Ce, Sr), and NaI(Tl) crystals were 0.22, 0.17, and 0.06, respectively. In contrast, Compton electron testing resulted in NLSD values of 0.11, 0.03, and 0.06 for the same crystals. The non-linear curves of these domestic crystals exhibited different slopes (Fig.~\ref{fig:LaBr3(Ce,Sr)}, Fig.~\ref{fig:LaBr3(Ce)} and Fig.~\ref{fig:NaI(Tl)}), indicating varying degrees of non-linearity at low energies. Based on the experimental results, the non-linearity of the three crystals to X/$\gamma$-rays exceeds that of Compton electrons, which can be attributed to the distinct interaction mechanisms between the incident particles and the material. 

The NLSD values for LaBr$_3$(Ce) were 1.29 times higher for X-rays and 3.67 times higher for Compton electrons compared to LaBr$_3$(Ce, Sr), indicating that the LaBr$_3$(Ce, Sr) crystal exhibited better linearity and suggesting that doping with Sr$^{2+}$ ions could improve non-linearity. However, the absolute light yield of the LaBr$_3$(Ce, Sr) crystal was slightly lower than that of LaBr$_3$(Ce) (Table~\ref{tab:absolute light yield}), potentially owing to the need for further optimization of the growth process and doping ratio by domestic manufacturers. The energy resolution of our LaBr$_3$(Ce, Sr) crystal was inferior to that reported by its foreign counterparts\cite{2013Improvement}. This discrepancy may arise from inherent performance variations among different crystals, differences in measurement methods when coupled with PMT, or distinctions in growth processes and raw materials between Chinese and Saint-Gobain crystals.

NaI(Tl) crystal exhibited "excess" light output of up to 9.2\% when tested with X-rays and 15.5\% when tested with Compton electrons. This "excess" light output positioned NaI(Tl) crystal as a distinct advantage for detecting low-energy X/$\gamma$-rays. The calibration and in-orbit performance of GECAM-C have validated that the NaI(Tl) crystals exceeded expectations\cite{ZHANG2023168586,Ground-calibration-GECAM-C}. In the mission of gamma-ray burst detection, energy resolution is not the primary concern. While NaI(Tl) crystals may not match the energy resolution and absolute light yield of LaBr$_3$ crystals, test results have demonstrated their satisfactory performance in the energy range of 10 keV to 1000 keV. Furthermore, NaI(Tl) crystals can be manufactured in larger sizes and are cost-effective. Consequently, GECAM-D utilized NaI(Tl) crystals as the sensitive detector materials.

We conducted a study on the light yield and non-linearity of the three crystals produced by the Beijing Glass Research Institute. An important insight from this study is that different calibration standards are required for the detection of gamma-rays and electrons. While the current GECAM satellite's GRDs lack electron-gamma discrimination capabilities, the non-linearity results of Compton electrons may find application in future corrections for electron detection.



\section{Acknowledgments}

This research was supported by the National Key Research and Development Program (Grant Nos. 2022YFB3503600 and 2021YFA0718500) and the Strategic Priority Research Program of the Chinese Academy of Sciences (Grant No. XDA15360102), and National Natural Science Foundation of China (Grant Nos. 12273042 and 12075258).

\bibliography{mybibfile}

\end{document}